\documentclass[12pt]{article} \usepackage{dina4p} \usepackage{my}
\usepackage{epsfig,amsmath} \usepackage{graphics}
 
\def\lsim{\mathrel{\rlap{\lower4pt\hbox{\hskip1pt$\sim$}}
    \raise1pt\hbox{$<$}}}                
\def\gsim{\mathrel{\rlap{\lower4pt\hbox{\hskip1pt$\sim$}}
    \raise1pt\hbox{$>$}}}                
\newcommand{\cA}{{\cal A}}

\newcommand{\tqn}{{q}_{t\;n}}
\newcommand{\Pmax}{\bar{q}}
\newcommand{\kt}{k_{t}}

\newcommand{\SMALLXC}{SMALLXa,SMALLXb}
\newcommand{\CCFM}{CCFMa,CCFMb,CCFMc,CCFMd}

\newcommand{\alphasb}{\bar{\alpha}_s}
\newcommand{\JETSETMC}{\PYTHIAMC}
\newcommand{\LEPTOMC}{Ingelman_LEPTO65}
\newcommand{\PYTHIAMC}{Jetsetc}

\def\CASCADE{{\sc Cascade}}
\def\SMALLX{{\sc Smallx}}

\def\LEPTO{{\sc Lepto}}
\def\PYTHIA{{\sc Pythia}}
\def\JETSET{{\sc Jetset}}
\makeindex
\begin{document}
%
\begin{flushright}
 DESY 01-114\\
 LUNFD6/(NFFL--7202) 2001 \\
 September 2001
\end{flushright}
\begin{center}  \begin{bf}{ \Large
The CCFM Monte Carlo generator  {\boldmath  \sc Cascade}}
  \end{bf} \\ \vspace{0.5cm}
{ \Large H. Jung }\\
    \vspace{0.5cm} {\large University of Lund, Department of Physics, Sweden }
\end{center}
\begin{abstract}
\CASCADE\ is a full hadron level Monte Carlo event generator for $ep$, 
$\gamma p$ and $p\bar{p}$ processes,
 which uses the
CCFM evolution equation for the initial state cascade in a backward evolution
approach supplemented 
with off - shell matrix elements for the hard scattering.
A detailed program description is given, with emphasis on parameters
the user wants to change and common block variables which 
completely specify the generated events.
\end{abstract}

\section{Tabular Summary}
\begin{tabular}{l l}
 program name & \CASCADE \\
 version      & 1.00/01    \\
 date of latest version& July 2001\\
 author      &  Hannes Jung (Hannes.Jung@desy.de)\\
 program size & $\sim$ 8000 lines of code\\
 input files needed & ccfm.dat (kms.dat, kmr.dat)         \\
 computer types & any with standard Fortran 77, tested on 
                 SGI, HP-UX, SUN, PC\\
 operating systems & Unix, Linux\\
 applicability & Deep Inelastic $ep$ Scattering \\
               & Photo-production in $ep$ Scattering \\
               & $\gamma p$ Scattering \\
               & heavy quark production in $pp$ and $p\bar{p}$ Scattering \\	   
  hard sub-processes included
  &  $  \gamma^* g^* \rightarrow q \bar{q} $, $Q\bar{Q}$\\  
  &  $  \gamma g^* \rightarrow J/\psi g $\\
  &  $  g^* g^* \rightarrow q \bar{q} $, $Q\bar{Q}$\\
 QCD cascade & initial state parton shower according to CCFM  \\
             & final state parton shower with angular ordering \\
 fragmentation model & LUND string \\
other programs called & \PYTHIA\  6.1 \\
                       & {\sc Bases/Spring} 5.1 \\
availability  & 
\verb+http://www.quark.lu.se/~hannes/cascade/+ \\
\end{tabular}
\newpage

\section{The CCFM evolution equation}
\label{sec:CCFMEquation}

The formulation of the CCFM~\cite{\CCFM} parton evolution 
for the implementation into a full hadron level Monte Carlo program is
described in detail
in~\cite{CASCADE,Jung_Salam_2000}.  Here only 
the main results are summarized and discussed for the case of leptoproduction.
\begin{figure}
\begin{center} 
\begin{picture}(0,0)%
\epsfig{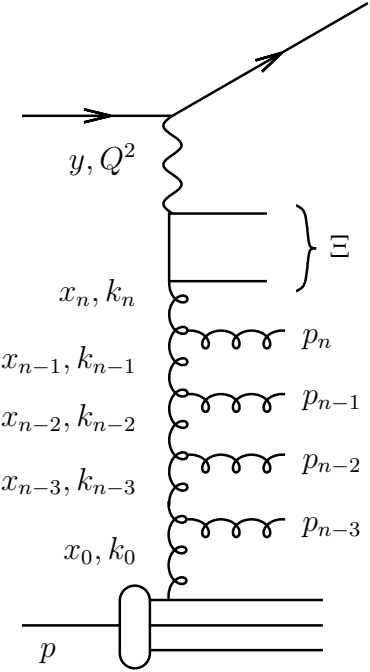}%
\end{picture}%
\setlength{\unitlength}{4144sp}%
\begingroup\makeatletter\ifx\SetFigFont\undefined%
\gdef\SetFigFont#1#2#3#4#5{%
  \reset@font\fontsize{#1}{#2pt}%
  \fontfamily{#3}\fontseries{#4}\fontshape{#5}%
  \selectfont}%
\fi\endgroup%
\begin{picture}(2959,4027)(4083,-3841)
\put(5637,-826){\makebox(0,0)[rb]{\smash{\SetFigFont{12}{14.4}
{\familydefault}{\mddefault}{\updefault}$y,Q^2$}}}
\put(5637,-1618){\makebox(0,0)[rb]{\smash{\SetFigFont{12}{14.4}
{\familydefault}{\mddefault}{\updefault}$x_n,k_n$}}}
\put(5637,-3163){\makebox(0,0)[rb]{\smash{\SetFigFont{12}{14.4}
{\familydefault}{\mddefault}{\updefault}$x_0,k_0$}}}
\put(5637,-2017){\makebox(0,0)[rb]{\smash{\SetFigFont{12}{14.4}
{\familydefault}{\mddefault}{\updefault}$x_{n-1},k_{n-1}$}}}
\put(5637,-2366){\makebox(0,0)[rb]{\smash{\SetFigFont{12}{14.4}
{\familydefault}{\mddefault}{\updefault}$x_{n-2},k_{n-2}$}}}
\put(5637,-2746){\makebox(0,0)[rb]{\smash{\SetFigFont{12}{14.4}
{\familydefault}{\mddefault}{\updefault}$x_{n-3},k_{n-3}$}}}
\put(6631,-1868){\makebox(0,0)[lb]{\smash{\SetFigFont{12}{14.4}
{\familydefault}{\mddefault}{\updefault}$p_n$}}}
\put(6631,-2235){\makebox(0,0)[lb]{\smash{\SetFigFont{12}{14.4}
{\familydefault}{\mddefault}{\updefault}$p_{n-1}$}}}
\put(6631,-2615){\makebox(0,0)[lb]{\smash{\SetFigFont{12}{14.4}
{\familydefault}{\mddefault}{\updefault}$p_{n-2}$}}}
\put(6631,-2995){\makebox(0,0)[lb]{\smash{\SetFigFont{12}{14.4}
{\familydefault}{\mddefault}{\updefault}$p_{n-3}$}}}
\put(6792,-1351){\makebox(0,0)[lb]{\smash{\SetFigFont{12}{14.4}
{\familydefault}{\mddefault}{\updefault}$\Xi$}}}
\put(5062,-3747){\makebox(0,0)[lb]{\smash{\SetFigFont{12}{14.4}
{\familydefault}{\mddefault}{\updefault}$p$}}}
\end{picture}
\end{center} 
\caption{{\it Kinematic variables for multi-gluon emission. The $t$-channel gluon
four-vectors are given by $k_i$ and the gluons emitted in the initial state
cascade have four-vectors $p_i$. The maximum angle for any emission is obtained
from the quark box, as indicated with $\Xi$.  
\label{CCFM_variables} }} 
\end{figure}

Figure~\ref{CCFM_variables} shows the pattern of QCD initial-state
radiation in a small-$x$ DIS event, together with labels for the
kinematics.  According to the CCFM evolution equation, the emission of
partons during the initial cascade is only allowed in an
angular-ordered region of phase space. 
The maximum allowed angle $\Xi$ for any gluon emission sets the scale 
$\Pmax$ for the
evolution and  
is defined by the hard scattering quark box,
 which connects the exchanged gluon to the virtual photon. In terms
of Sudakov variables the quark pair momentum is written as:
\begin{equation}
p_q + p_{\bar{q}} = Y (p_p + \Xi p_e) + Q_t
\end{equation}
where $p_p$ and $p_e$ are the proton and electron momenta,
respectively and $Q_t$ is the transverse momentum of the quark pair and 
$Y$ is its
light-cone momentum fraction. The scale $\Pmax$ is related to the maximum angle 
$\Xi$ via $\Pmax = x_n \sqrt{\Xi s}$, with $s$ being the squared 
center of mass energy $s=(p_e+p_p)^2$.
\par
The CCFM evolution equation can be written in a differential form~\cite{CCFMd},
which is best suited for a backward evolution approach adopted in the Monte
Carlo generator \CASCADE\ ~\cite{CASCADE,Jung_Salam_2000}:
\begin{equation}
\Pmax^2\frac{d\; }{d \Pmax^2} 
   \frac{x \cA(x,\kt,\Pmax)}{\Delta_s(\Pmax,Q_0)}=
   \int dz \frac{d\phi}{2\pi}\,
   \frac{\tilde{P} (z,\Pmax/z,\kt)}{\Delta_s(\Pmax,Q_0)}\,
 x'\cA(x',\kt',\Pmax/z) 
\label{CCFM_differential}
\end{equation} 
where $\cA(x,\kt,\Pmax)$ is the unintegrated gluon density, depending on 
$x$, $\kt$ and the evolution variable $\Pmax$. The splitting variable is 
$z=x/x'$ and $\vec{\kt}' = (1-z)/z\vec{q} + \vec{\kt}$, where the vector
$\vec{q}$ is at an azimuthal angle $\phi$.
The  Sudakov form factor $\Delta_s$ is given by:
\begin{equation}
\Delta_s(\Pmax,Q_0) =\exp{\left(
 - \int_{Q_0^2} ^{\Pmax^2}
 \frac{d q^{2}}{q^{2}} 
 \int_0^{1-Q_0/q} dz \frac{\alphasb(q^2(1-z)^2)}{1-z}
  \right)}
  \label{Sudakov}
\end{equation}
with $\alphasb=\frac{C_A \alpha_s}{\pi}=\frac{3 \alpha_s}{\pi}$. For
inclusive quantities at leading-logarithmic order the Sudakov form
factor cancels against the $1/(1-z)$ collinear singularity of the
splitting function. 
\par
The splitting function $\tilde{P}_g (z_i,q_i,k_{ti})$ for branching $i$ 
is given by:
\begin{equation}
\tilde{P}_g (z_i,q_i,k_{ti})
= \frac{\alphasb(q^2_{i}(1-z_i)^2)}{1-z_i} + 
\frac{\alphasb(k^2_{ti})}{z_i} \Delta_{ns}(z_i,q^2_{i},k^2_{ti})
\label{Pgg}
\end{equation}
where the non-Sudakov form factor $\Delta_{ns}$ is defined as:
\begin{equation}
\log\Delta_{ns} =  -\alphasb(k^2_{ti})
                  \int_0^1 \frac{dz'}{z'} 
                        \int \frac{d q^2}{q^2} 
              \Theta(k_{ti}-q)\Theta(q-z'q_{ti})
                  \label{non_sudakov}                   
\end{equation}
\par
The splitting function $\tilde{P} (z,q,\kt)$ in eq.(\ref{Pgg})
contains only the singular parts in $z$ and $(1-z)$. The finite terms in 
the splitting function are neglected, since they
are not obtained in CCFM at the leading infrared
accuracy~\cite[p.72]{CCFMc}. As soon as they become available they can be
easily implemented into the Monte Carlo generator.
\section{Backward evolution: CCFM and {\boldmath\CASCADE }}
\label{sec:Backward}

The idea of a backward evolution~\cite{PYTHIAPSa,LEPTOPS}
is to first generate the hard
scattering process with the initial parton momenta distributed
according to the parton distribution functions.  This involves in
general only a fixed number of degrees of freedom, and the hard
scattering process can be generated quite efficiently. The
initial state cascade is generated by going backwards from the hard
scattering process towards the beam particles. 

According to the CCFM equation the probability of finding a gluon in
the proton depends on three variables, the momentum fraction $x$, the
transverse momentum squared $k_t^2$ of the exchanged gluons and the
scale $\Pmax = x_{n} \sqrt{s \Xi}$, which is related to the
maximum angle allowed for any emission 
$\Xi$. To solve eq.(~\ref{CCFM_differential}) the unintegrated gluon
distribution $\cA(x,k,\Pmax)$ has to be determined beforehand. 
\par
Given this distribution, the generation of a full hadronic event is separated
into three steps, as implemented in the hadron-level Monte Carlo program 
\CASCADE: 
\begin{itemize}
\item[$\bullet$] 
The hard scattering process is generated,
\begin{equation}
\sigma = \int dk_t^2 dx_g {\cal A}(x_g,k_t,\Pmax)
 \sigma (\gamma^* g^* \to X )\,,
\label{x_section}
\end{equation}
with $X$ being $ q\bar{q}$, $ Q\bar{Q}$ or $J/\psi g$ states. The hard cross
section is calculated 
using the off-shell matrix elements given in~\cite[p.~178
ff]{CCH} for $ q\bar{q}$ and $ Q\bar{Q}$ or for the case of
$\gamma g^* \to J/\psi g$ in~\cite{saleev_zotov_a}.
The gluon momentum is given in Sudakov representation:
\begin{equation}
 k = x_g p_p + \bar{x}_g p_e + k_t \simeq x_g p_p  + k_t\,.
 \label{xgluon}
\end{equation}
where the last expression comes from the high energy approximation ($x_g \ll
1$), which then gives  $-k^2 \simeq k_t^2$.
\item[$\bullet$] The initial state cascade is generated according to
  CCFM in a backward evolution approach.
\item[$\bullet$] The hadronization is performed using the Lund string
  fragmentation implemented in \PYTHIA\ /\JETSET \cite{\JETSETMC}.
\end{itemize}

The backward evolution there faces one difficulty: The gluon
virtuality enters in the hard scattering process and also influences
the kinematics of the produced quarks and therefore the maximum angle
allowed for any further emission in the initial state cascade. This
virtuality is only known after the whole cascade has been generated,
since it depends on the history of the gluon evolution
(as $\bar{x}_g$ in eq.(~\ref{xgluon}) may not be neglected for exact
kinematics).  In the
evolution equations itself it does not enter, since there only the
longitudinal energy fractions $z$ and the transverse momenta are
involved.  This problem can only approximately be overcome by using
$k^2 = k_t^2/(1-x_g)$ for the virtuality which is correct in the case
of no further gluon emission in the initial state.

The Monte Carlo program \CASCADE~ can be used to generate unweighted
full hadron-level events, including initial-state parton evolution
according to the CCFM equation and the off-shell matrix elements for
the hard scattering process. It is applicable for $p \bar{p}$, 
photo-production as well as for deep inelastic scattering.
  The typical
time needed to generate one event is $\sim 0.03$~sec on a Pentium II 
(266~MHz), which is
similar to the time needed by standard Monte Carlo event generators
such as \LEPTO~\cite{\LEPTOMC} or \PYTHIA~\cite{\PYTHIAMC}.  

\subsection{The unintegrated gluon density}
The CCFM unintegrated gluon density $x {\cal A}(x,k_{t},\Pmax)$ is
obtained from a forward evolution procedure as implemented in
\SMALLX~\cite{\SMALLXC} by a fit to the measured structure function $F_2$ as
described in~\cite{CASCADE,Jung_Salam_2000}. 
From the initial gluon distribution  a set of values $x$
and $k_{t}$ are obtained by evolving up to a given scale $\log
\Pmax$ using the forward evolution procedure.  This is
repeated $10^8$ times thus obtaining a distribution of the
unintegrated gluon density $x {\cal A}(x,k_{t\;},\Pmax)$ for the
slice of phase space with a given $\Pmax$ ($\Pmax > \tqn$). To obtain
a distribution in $\log \Pmax$, the above procedure is repeated from
the beginning $50$ times for the different grid points in $\log \Pmax$
up to $\Pmax = 1800$~GeV. 
Due to the complicated structure of the CCFM
equation, no attempt is made to parameterize the unintegrated gluon
density. Instead, the gluon density is calculated on a grid in $\log
x$, $\log k_{t}$ and $\log \Pmax$ with $50 \times 50 \times 50$ points
and a linear interpolation is used to obtain the gluon density for
values in between the grid points.
The data file (\verb+ccfm.dat+) containing the  
$50 \times 50 \times 50$ grid points 
is read in at the beginning of the program.
\par
The parameter \verb+IGLU+ offers the possibility to compare 
the CCFM unintegrated gluon density 
$x {\cal A}(x,k_{t},\Pmax)$ (\verb+IGLU=1+)
 with other unintegrated gluon densities
published:
a simple numerical derivative of a standard integrated gluon density
$\frac{d xg(x,Q^2)}{dQ^2}$ taken
from~\cite{GRV95}  (\verb+IGLU=2+), the one in the
approach 
of Bl\"umlein~\cite{Bluemlein} and coded 
by~\cite{baranov_zotov_1999,baranov_zotov_2000} (\verb+IGLU=3+),
the unintegrated gluon density of KMS\footnote{A. Stasto kindly
provided the program.}~\cite{martin_stasto} 
(\verb+IGLU=4+, stored in \verb+kms.dat+), 
the one of the saturation model by~\cite{wuesthoff_golec-biernat}
 (\verb+IGLU=5+) and the one of KMR\footnote{M. Kimber kindly
provided the program.}~\cite{martin_kimber} (\verb+IGLU=6+, 
stored in \verb+kmr.dat+).

Except for the default CCFM unintegrated gluon density with (\verb+IGLU=1+), 
no
initial state parton shower can be generated, because the angular variable,
essential for angular ordering in the initial state cascade, is not present.
However, the transverse momenta of the incoming partons are properly treated.
\subsection{$\alpha_s$ and the choice of scales}
The strong coupling $\alpha_s$ is calculated via the \PYTHIA~\cite{\PYTHIAMC} 
subroutine \verb+PYALPS+. 
Maximal and minimal number of flavors used in $\alpha_s$ are set by
\verb+MSTU(113),MSTU(114)+, $\Lambda_{QCD}$ =\verb+PARU(112)+ with respect to
\verb+MSTU(112)+ flavors and stored in the \PYTHIA\ common 
\verb+COMMON/PYDAT1/+.
In the initial state cascade according to CCFM, the transverse momenta of the
$t$-channel gluons are allowed to perform a random walk for small $z$ values 
and $k_t$ can become very small. 
In the $1/z$ part of the splitting function we use $\mu=k_t$ as the scale in 
$\alpha_s(\mu)$ and in the $1/(1-z)$ part  $\mu=(1-z) q_t$ is used. In
addition we require $\mu > Q_0=1.4$ GeV, resulting in 
$\alpha_s(\mu>Q_0) < 0.6$.
The parameter  $Q_0=1.4$ GeV was determined from the requirement to describe the
structure function $F_2$ as described in~\cite{CASCADE,Jung_Salam_2000}.
\par
The scale $\mu$ which is used in $\alpha_s$
in the hard scattering matrix element can be changed
with the parameter \verb+IQ2+, the default choice is $\mu^2 = p_t^2$.

\subsection{Quark masses}

The quark mass for light quarks ($u,d,s$)
 is fixed to $m_q=0.140$ GeV.
 This, together with the treatment of $\alpha_s$ at small scales
$\mu$, gives also a reasonable total cross-section for photo-production at HERA
energies. The masses for heavy quarks are given by the \JETSET\ / \PYTHIA\ 
defaults ($m_c=1.5$ GeV, $m_b=4.8$ GeV) and can be changed according to the
\PYTHIA\ prescription.

\subsection{Leptoproduction}  

\CASCADE\ can be used to simulate leptoproduction events over the whole $Q^2$
range. By fixing the light quark masses to  $m_q=0.140$ GeV
and $\alpha_s$ for small $\mu$, the hard scattering
matrix element remains finite over the full phase space.
The total cross section is simulated by selecting \verb+IPRO=10+ and 
\verb+NFLAV=4(5)+. With \verb+IPRO=10+ light quarks ($u,d,s$) are selected  and
with \verb+NFLAV>3+ the program automatically includes heavy flavor production
via the process \verb+IPRO=11+ and \verb+IHFLA=4+ up to \verb+IHFLA=NFLAV+.
\par
Heavy flavor production can be generated separately via \verb+IPRO=11+. 
The value of \verb+IHFLA+ determines the heavy flavor to be generated.
\par
The matrix element for $\gamma g^* \to J/\psi g$ calculated in
\cite{saleev_zotov_a} is available for quasi-real $\gamma$'s via the process
\verb+IPRO=2+.

\subsection{Photoproduction}  

\CASCADE\ can be used to simulate 
real photoproduction events by using \verb+KE=22+. The same options as for
leptoproduction are available.

\subsection{Hadroproduction}

\CASCADE\ can be used to simulate heavy quark production in
 $p p$ or $p \bar{p}$ collisions (\verb+IPRO=14+ for heavy flavor production,
 and \verb+IHFLA=4(5)+ for charm (bottom) quarks).
 The flavor code (\verb+KE+) for beam 1 can be
 chosen as \verb+KE=2212+ for proton or \verb+KE=-2212+ for anti - proton,
 beam 2 is always a proton. 
 
\subsection{Initial and final state parton shower}

Initial state parton showers are generated in a backward evolution approach
described in detail in~\cite{CASCADE,Jung_Salam_2000}. The initial 
state parton shower
consists only of gluon branchings and is
generated in an angular ordered region in the laboratory frame. In the present
version, the gluons emitted during the branchings are treated on-mass shell,
and no further time - like branching occurs.
\par
All parameters (like the scale $\mu$
in $\alpha_s$, the collinear cut-off $Q_0$) for  the initial state cascade  
are fixed 
from the determination of the unintegrated gluon density. The
transverse momenta of the partons which enter the hard scattering 
matrix element
are already generated in the beginning and are not changed
during the whole initial and final state
parton showering (in contrast to standard DGLAP type parton shower Monte Carlo
generators like \PYTHIA\ ~\cite{\PYTHIAMC} or \LEPTO\ ~\cite{\LEPTOMC}). 
\par
The final state parton shower uses the parton shower routine \verb+PYSHOW+ 
of \PYTHIA\
with the scale \verb+QMAX=+$(m_{1\;\perp}+m_{2\;\perp})/2$,
with $m_{1(2)\;\perp}$ being the transverse mass of the hard parton 1(2).

\subsection{Remnant treatment}
The proton remnant is built in subroutine \verb+CAREMN+, which is a slightly
modified version of the \PYTHIA\ / \LEPTO\ subroutine \verb+PYREMN+. In the
present version no intrinsic transverse momentum, in addition  to the transverse
momentum from the initial state cascade, is included.

\section{Description of the program components}
In \CASCADE\ all variables are declared \verb"Double Precision". The 
Lund string model is used for hadronization as implemented in \PYTHIA\ 6.1 
\cite{\PYTHIAMC}. The final state QCD radiation is performed via 
\verb"PYSHOW" from
 \PYTHIA\ 6.1. 
 The treatment of the proton remnant follows very closely the ones in
 \LEPTO \cite{\LEPTOMC} for the leptoproduction case and the one in 
\PYTHIA\ 6.1 for the proton - proton case. However slight modifications were needed
to adapt to the cascade treatment here.
\par
The unintegrated gluon density is stored on data files
(\verb+ccfm.dat,kms.dat,kmr.dat+), 
and is read in at the beginning of the program.
\par
The program has to be loaded together with \PYTHIA\ 6, to ensure that the double
precision code of \JETSET\  is loaded.
\subsection{Random number generator}
Since the variables are declared in double precision, also a double precision
random number generator has to be used to avoid any bias. The function
\verb+DCASRN+ gives a single random number,  the function \verb+DCASRNV+ returns
an array of length \verb+LEN+ of random numbers. The default random number
generator is \verb+RM48+ (called in \verb+DCASRN+ and \verb+DCASRNV+) from 
{\sc Cernlib}. The user can change this to any preferred \verb+Double Precision+
random number generator.
\subsection{Integration and event generation}
The integration of the total cross section and the generation of unweighted
events is performed with the help of {\sc Bases/Spring} \cite{bases}.

\subsection{Subroutines and functions}

The source code of \CASCADE\ and this manual can be found under:\\
\verb+http://www.quark.lu.se/~hannes/cascade/+

\begin{defl}{123456789012345}
\item[{\tt CAMAIN}]
                  main program.
\item[{\tt CASINI}] 
                   to initializes the program.
\item[{\tt CASCADE}]
     to perform integration of the cross section. This routine has to be
            called before event generation can start.
\item[{\tt CAEND }] 
           to print the cross section and the number of events.
\item[{\tt CAUNIGLU(KF,X,KT,P,XPQ) }]   
      to extract the unintegrated gluon density 
	$x {\cal A}(x,k_{t},\Pmax)$ for a proton with \verb+KF=2212+,
	as a function of $x=$\verb+X+, $k_{t}^2=$\verb+KT+ and $\Pmax=$\verb+P+.
      The gluon density is returned in \verb+XPQ(0)+, where \verb+XPQ+ is an 
      array with \verb+XPQ(-6:6)+.
\item[{\tt EVENT }] 
        to perform the event generation.
\item[{\tt ALPHAS(RQ)}] 
         to give $\alpha_s (\mu)$ with $\mu = $\verb+RQ+.
\item[{\tt PARTI }] 
         to give initial particle and parton momenta.
\item[{\tt FXN1 }] 
         to call routines for selected processes:
          \verb"XSEC1".
\item[{\tt CUTG(IPRO) }] 
          to cut on $p_t$ for $2 \to 2$ process 
          in integration and event generation.
\item[{\tt MEOFFSH }] 
          matrix element for 
	    $\gamma^* g^* \rightarrow q \bar{q}$  and
	    $g^* g^* \rightarrow q \bar{q}$ 
	    including masses. $q$ can be light or heavy quarks.
\item[{\tt DOT(A,B) }] 
         four-vector dot product of $A$ and $B$.
\item[{\tt DOT1(I,J)}]
          four-vector dot product of vectors I and J in
          \verb"PYJETS" common.
\item[{\tt PHASE }] 
         to generate momenta of final
         partons in a $2 \rightarrow 2$ subprocess according to phase space	   
\item[{\tt P\_SEMIH }] 
         to generate kinematics and the 
         event record for $ep$, $\gamma p$ and $p\bar{p}$
         processes.
\item[{\tt CAREMN(IPU1,IPU2) }]   
         to generate the beam remnants.
            Copied from LEPTO 6.1~\cite{\LEPTOMC} and updated for
            the use in \CASCADE . 
\item[{\tt CASPLI(KF,KPA,KFSP,KFCH) }]   
            to give the spectator \verb"KFSP" and \verb"KFCH"
            partons when a parton
            \verb"KPA" is removed from particle \verb"KF".
            Copied from LEPTO 6.1~\cite{\LEPTOMC} and updated for
            the use \CASCADE. 
\item[{\tt CAPS }]   
  to generate color flow  for all
            processes and prepare for initial and final state
            parton showers.
\item[{\tt CASCPS(IPU1,IPU2) }]   to generate initial state radiation.
\item[{\tt GADAP }]   Gaussian integration routine for 1-dim and 
            2-dim integration.
            Copied from LEPTO 6.1~\cite{\LEPTOMC}.
\end{defl}

\subsection{Parameter switches}
\begin{defl}{123456789012345}
\item[]   BASES/SPRING Integration procedure.
\item[{\tt NCAL:}]  (D:=20000) Nr of calls per iteration for bases.
\item[{\tt ACC1:}]  (D:=1)    relative precision (in \%) for grid optimization.
\item[{\tt ACC2:}]  (D:=0.5)  relative precision (in \%) for integration.
\end{defl}

\subsubsection{Parameters for kinematics}
\begin{defl}{123456789012345}
\item[{\tt PLEPIN:}] \index{PLEPIN}
                     (D:=$-30$)   momentum $p$ [GeV/$c$]
                        of incoming electron (\verb"/INPU/").
\item[{\tt PIN:}] \index{PIN}
                     (D:=$820$    momentum $p$ [GeV/$c$]
                        of incoming proton (\verb"/INPU/").
\item[{\tt QMI:}] \index{QMI} 
       (D: = 5.0) (\verb"/VALUES/") minimum $Q^2$ to be generated.
\item[{\tt QMA:}] \index{QMA}
 (D: = $10^8$) (\verb"/VALUES/") maximum $Q^2$ to be generated.
\item[{\tt YMI:}] \index{YMI} 
(D: = 0.0)(\verb"/VALUES/") minimum $y$ to be generated.
\item[{\tt YMA:}] \index{YMA}  
(D: = 1.0) (\verb"/VALUES/") maximum $y$ to be generated.
\item[{\tt THEMA,THEMI}] \index{THEMA,THEMI}
  (D: {\tt THEMA} = 180., {\tt THEMI} = 0)
                    maximum and minimum scattering angle $\theta$ of the 
                    electron ({\tt /CAELEC/}).
\item[{\tt NFLAV}]  \index{NFLAV}
(D: = 5) number of active flavors, can be set by user 
                    ({\tt /CALUCO/}).

\end{defl}

\subsubsection{Parameters for hard subprocess selection}
\begin{defl}{123456789012345}
\item[{\tt IPRO:}] \index{IPRO} (D: = 10) ({\tt /CAPAR1/})
                    selects hard subprocess to be generated. 
\item[{\it         }]
                 =2:   $\gamma g^* \rightarrow J/\psi g$
\item[{\it         }]
                 =10:   $\gamma^* g^* \rightarrow q \bar{q}$
                        for light quarks.
\item[{\it         }]
                 =11:   $\gamma^* g^* \rightarrow Q \bar{Q}$
                        for heavy quarks.

\item[{\tt PT2CUT(IPRO):}] \index{PT2CUT}
 (D=0.0) minimum $\hat{p}^2 _{\perp}$ for
                            process {\tt IPRO} ({\tt /CAPTCUT/}).       
\end{defl}
\subsubsection{Parameters for parton shower and fragmentation}
\begin{defl}{123456789012345}
\item[{\tt NFRAG:}] \index{NFRAG} (D: = 1)
                        switch for fragmentation({\tt /CAINPU/}).
\item[] = 0 off
\item[] = 1 on 
\item[{\tt IFPS:}] \index{IFPS} (D: = 3)
                  switch  parton shower({\tt /CAINPU/}).
\item[] = 0 off
\item[] = 1 initial state
\item[] = 2 final state
\item[] = 3 initial and final state  
\item[{\tt ICCFM:}] \index{ICCFM} (D: =1)
\item[] =1 CCFM evolution.
\end{defl}

\subsubsection{Parameters for structure functions $\alpha_s$ and scales}
\begin{defl}{123456789012345}
\item[{\tt IRUNAEM:}] \index{IRUNAEM} (D: = 0) ({\tt /CAPAR1/})
                   select running of $\alpha _{em}(Q^2)$.
\item[]
                        =0:  no running of $\alpha _{em}(Q^2)$
\item[]
                        =1:  running of $\alpha _{em}(Q^2)$

\item[{\tt IRUNA:}] \index{IRUNA} (D: = 1)
                        switch for running $\alpha _s$.
\item[]
                        =0:  fixed $\alpha_s=0.3$ 
\item[]
                        =1: running $\alpha _s(\mu^2)$
\item[{\tt IQ2:}] \index{IQ2} (D: = 3)
                   select scale $\mu^2$ for $\alpha _s(\mu^2)$.
\item[]
                        =1:  $\mu^2 = 4 \cdot m_{q} ^2$
                             (use only for heavy quarks!)
\item[]
                        =2:  $\mu^2 = \hat{s} $
                             (use only for heavy quarks!)
\item[]
                        =3:  $\mu^2 = 4 \cdot m^2 + p_{\perp} ^2$
                             
\item[]
                        =4:  $\mu^2 = Q^2$
                        
\item[]
                        =5:  $\mu^2 = Q^2 + p_{\perp} ^2$
\item[{\tt IGLU:}] \index{IGLU} (D: = 1)
			select unintegrated gluon density{\tt /GLUON/}).
\item[]	Note for \verb+IGLU+$>1$, all initial state parton showers are
switched off.		
\item[]
                        =1:  CCFM unintegrated gluon	(\verb+ccfm.dat+).	
\item[]
                        =2:  derivative of GRV~\cite{GRV95}
				$\frac{d xg(x,Q^2)}{dQ^2}$.
\item[]
                        =3:  approach of Bl\"umlein~\cite{Bluemlein}.
\item[]
                        =4:  KMS~\cite{martin_stasto} (\verb+kms.dat+).
\item[]
                        =5:  saturation model~\cite{wuesthoff_golec-biernat}.
\item[]
                        =6:  KMR~\cite{martin_kimber} (\verb+kmr.dat+).

\end{defl}

\subsubsection{Accessing information}
\begin{defl}{123456789012345}
\item[ ]

\item[{\tt AVGI}] \index{AVGI} integrated cross section ({\tt /CAEFFIC/}).
\item[{\tt SD}] \index{SD} standard deviation of integrated cross section
                 ({\tt /CAEFFIC/}).
\item[ ]
\item[{\tt SSS}] \index{SSS} 
squared center of mass energy  $s$ ({\tt /CAPARTON/}).
\item[{\tt PBEAM}] \index{PBEAM} energy momentum vector of beam particles
  ({\tt /CABEAM/}).
\item[{\tt KBEAM}] \index{KBEAM} flavor code of beam particles
 ({\tt /CABEAM/}).
\item[{\tt Q2}] \index{Q2} in leptoproduction: actual $Q^2$ of
                         $\gamma$ ({\tt /CAPAR4/}).
\item[{\tt YY}] \index{YY}
negative light-cone momentum
                       fraction of parton $1$ ($\gamma^*$, $g^*$)
				({\tt /CASGKI/}).
\item[{\tt YY\_BAR}] \index{YY\_BAR}
positive light-cone momentum
                       fraction parton $1$ ($\gamma^*$, $g^*$)
				({\tt /CASGKI/}).
\item[{\tt XG}] \index{XG}
positive light-cone momentum
                        fraction of parton $2$ ($g^*$) 
				({\tt /CASGKI/}).
\item[{\tt XG\_BAR}] \index{XG\_BAR}
negative light-cone momentum
                        fraction of parton $2$ ($g^*$)
				({\tt /CASGKI/}).
\item[{\tt KT2\_1,KT2\_2}] \index{KT2\_1,KT2\_2} transverse momenta squared
			$k_{t\;1(2)}^2$ [GeV$^2$] of partons 
 			$1(2)$ which enter to the matrix element.
\item[{\tt YMAX,YMIN}] \index{YMAX,YMIN} actual upper and lower limits for
			$y=$\verb+YY+ 
                          ({\tt /CAPAR5/}).
\item[{\tt Q2MAX,Q2MIN}] \index{Q2MAX,Q2MIN} actual upper and
                        lower limits for $Q^2$ (corresponding to \verb+KT2_1+)
				 of $\gamma$ ({\tt /CAPAR5/}).
\item[{\tt XMAX,XMIN}] \index{XMAX,XMIN} 
upper and lower limits for $x$ ({\tt /CAPAR5/}).

\item[{\tt AM(18)}] \index{AM}
 vector of masses of final state particles of hard
                         interaction ({\tt /CAPAR3/}).
\item[{\tt SHAT}] \index{SHAT}
                        invariant mass $\hat{s}$ [GeV$^2$]
                     of hard subprocess ({\tt /CAPAR5/}).

\item[{\tt NIA1,NIA2}] \index{NIA1,NIA2}
 position of partons in hard interaction in
                         {\tt PYJETS} event record ({\tt /CAHARD/}).
\item[{\tt NF1,NF2}] \index{NF1,NF2} first and last position final
                         partons/particles of
                         hard interaction in {\tt PYJETS} ({\tt /CAHARD/}).
\item[{\tt Q2Q}] \index{Q2Q} hard scattering scale $\mu ^2$ used in
                         $\alpha_s$ and structure functions ({\tt /CAPAR4/}).
\item[{\tt ALPHS}] \index{ALPHS} actual $\alpha_s$ ({\tt /CAPAR2/}).
\item[{\tt ALPH}] \index{ALPH} $\alpha_{em}$ ({\tt /CAPAR2/}).
\item[{\tt NIN}] \index{NIN} 
number of trials for event generation ({\tt /CAEFFIC/}).
\item[{\tt NOUT}] \index{NOUT} number of successful generated events 
({\tt /CAEFFIC/}).

\end{defl}

\subsection{List of COMMON blocks}
\verb"  COMMON/CABEAM/PBEAM(2,5),KBEAM(2,5),KINT(2,5)" \\
\verb"  COMMON/CAHARD/ NIA1,NIA2,NIR2,NF1,NF2"\\
\verb"  COMMON/CAHFLAV/ IHFLA"\\
\verb"  COMMON/CAINPU/PLEPIN,PPIN,NFRAG,ILEPTO,IFPS,IHF,INTER,ISEMIH"\\
\verb"  COMMON/CALUCO/KE,KP,KEB,KPH,KGL,KPA,NFLAV"\\
\verb"  COMMON/CAEFFIC/AVGI,SD,NIN,NOUT"\\
\verb"  COMMON/CAELEC/THEMA,THEMI"\\
\verb"  COMMON/CAGLUON/IGLU"\\
\verb"  COMMON/CAPAR1/IPRO,IRUNA,IQ2,IRUNAEM"\\
\verb"  COMMON/CAPAR2/ALPHS,PI,ALPH,IWEI"\\
\verb"  COMMON/CAPAR3/AM(18),PCM(4,18)"\\
\verb"  COMMON/CAPAR4/Q2,Q2Q"\\
\verb"  COMMON/CAPAR5/SHAT,YMAX,YMIN,Q2MAX,Q2MIN,XMAX,XMIN"\\
\verb"  COMMON/CAPAR6/LST(30),IRES(2)"\\
\verb"  COMMON/CAPARTON/SSS,CM(4),DBCMS(4)"\\
\verb"  COMMON/CAPTCUT/PT2CUT(20)"\\
\verb"  COMMON/CASKIN/YY,YY_BAR,XG,XG_BAR,KT2_1,KT2_2,PT2H,SHH"\\
\verb"  COMMON/VALUES/QMI,YMI,QMA,YMA"\\
\section{Example Program}
\begin{verbatim}

      PROGRAM CASMAIN
      Implicit None
      Integer N1,N2
      REAL PLEPIN,PPIN
      INTEGER KE,KP,KEB,KPH,KGL,KPA,NFRAG,ILEPTO,IFPS,IHF
      INTEGER INTER,ISEMIH
      INTEGER NIA1,NIR1,NIA2,NIR2,NF1,NF2,NFT,NFLAV
      COMMON/CALUCO/KE,KP,KEB,KPH,KGL,KPA,NFLAV
      COMMON/CAINPU/PLEPIN,PPIN,NFRAG,ILEPTO,IFPS,IHF,INTER,ISEMIH
      COMMON/CAHARD/ NIA1,NIA2,NIR2,NF1,NF2
      INTEGER IHFLA
      COMMON/CAHFLAV/ IHFLA

      DOUBLE PRECISION THEMA,THEMI,PT2CUT
      INTEGER IRUNA,IQ2,IRUNAEM
      INTEGER IPRO
      COMMON/CAPAR1/IPRO,IRUNA,IQ2,IRUNAEM
      COMMON/CAELEC/ THEMA,THEMI
      COMMON/CAPTCUT/PT2CUT(20)
      REAL ULALPS,ULALEM
      EXTERNAL ULALPS,ULALEM
      DOUBLE PRECISION QMI,YMI,QMA,YMA
      COMMON/VALUES/QMI,YMI,QMA,YMA

      Integer Iglu
      Common/CAGLUON/Iglu
	
      Integer ISEED,I
	
	
      ISEED = 124567
      n1=0
      n2=0
C initialize random number generator	
      CALL RM48IN(ISEED,N1,N2)
C initialize PYTHIA 6 parameters
      CALL GPYINI
C initialize CASCADE parameters
      CALL CASINI


C Select parton shower (IPS=1 initial, =2 final, 3 initial+final PS )
      IFPS = 3
C scale for alpha_s
C IQ2 =1 mu^2 = m_q^2 (m_q = light quark or heavy quark depending on IPRO)
C IQ2 =2 mu^2  = shat
C IQ2 =3 mu^2  = m_q^2 + pt^2 (m_q = light quark or heavy quark depending on IPRO)
C IQ2 =4 mu^2  = q^2 (q^2 of virtual photon)
C IQ2 =5 mu^2  = q^2 + pt^2 (q^2 of virtual photon)
      IQ2=3
C select process (IPRO=10 for light quarks, IPRO=11 for heavy quarks)
      IPRO= 10
C total number of flavors involved
      NFLAV = 4
C select unintegrated gluon density (D=1)
      Iglu = 1
C minimum Q^2 of electron to be generated
      QMI = 0.5d0
C maximum Q^2 of electron to be generated
      QMA = 10D8
C minimum y of electron to be generated
      YMI=0.0d0
C minimum y of electron to be generated
      YMA=1.0d0
C maximum theta angle of scattered electron
      THEMA = 180.0D0
C minimum  theta angle of scattered electron
      THEMI =   0.0D0
C momentum of beam 1 (electron,proton,antiproton)
      PLEPIN =-27.5
C Lund flavor code for beam 1 (electron=11,photon=22,proton=2212,antiproton=-2212)
      KE=11
C momentum of beam 2 (proton)
      PPIN   = 820.
C perform fragmentation NFRAG=0/1
      NFRAG = 1
c for IPRO = 11  which flavor is produced
      IHFLA = 4
c 
c Start integration of x-section
c 
      CALL CASCADE
c 
c Print out result of integration of x-section
c 
      CALL CAEND(1)

c 
c Start event loop
c 
      Do I=1,100
c generate an event
         CALL EVENT
      Enddo
c 
c Print out of generated events summary
c 
      CALL CAEND(20)

      STOP
      END
\end{verbatim}

\section{Acknowledgments}
I am very grateful to B.~Webber for providing me with the \SMALLX~
code, which was the basis for the \CASCADE\ Monte Carlo generator.  
I am very grateful also to G.~Ingelman and T.~Sj\"ostrand for their courtesy to
let me use their code for proton remnant treatment.
I am also grateful to G.~Salam for all his patience and his help in all
different kinds of discussions concerning CCFM and a backward
evolution approach.
I have enjoyed and learned a lot from the discussions with
B.~Andersson, G.~Gustafson, L.~J\"onsson, H.~Kharraziha
and L.~L\"onnblad during several years. Many thanks also go to 
S.~Baranov and N.~Zotov who were the first raising my interest in 
$k_t$-factorization.
During all the years I had fun and great times with Antje.
 
\begin{theindex}

  \item ALPH, 9
  \item ALPHS, 9
  \item AM, 9
  \item AVGI, 9

  \indexspace

  \item ICCFM, 8
  \item IFPS, 8
  \item IGLU, 9
  \item IPRO, 8
  \item IQ2, 9
  \item IRUNA, 8
  \item IRUNAEM, 8

  \indexspace

  \item KBEAM, 9
  \item KT2\_1,KT2\_2, 9

  \indexspace

  \item NF1,NF2, 9
  \item NFLAV, 8
  \item NFRAG, 8
  \item NIA1,NIA2, 9
  \item NIN, 9
  \item NOUT, 9

  \indexspace

  \item PBEAM, 9
  \item PIN, 8
  \item PLEPIN, 8
  \item PT2CUT, 8

  \indexspace

  \item Q2, 9
  \item Q2MAX,Q2MIN, 9
  \item Q2Q, 9
  \item QMA, 8
  \item QMI, 8

  \indexspace

  \item SD, 9
  \item SHAT, 9
  \item SSS, 9

  \indexspace

  \item THEMA,THEMI, 8

  \indexspace

  \item XG, 9
  \item XG\_BAR, 9
  \item XMAX,XMIN, 9

  \indexspace

  \item YMA, 8
  \item YMAX,YMIN, 9
  \item YMI, 8
  \item YY, 9
  \item YY\_BAR, 9

\end{theindex}

\end{document}